\documentclass[a4paper,usenatbib,onecolumn]{mnras}
\usepackage[T1]{fontenc}
\usepackage{ae,aecompl}

\usepackage{graphicx}	% Including figure files
\usepackage{amsmath}	% Advanced maths commands
\usepackage{amssymb}	% Extra maths symbols

\newcommand{\beq}[0]{\begin{equation}}
\newcommand{\eeq}[0]{\end{equation}}

\title[Relativistic Turbulence with Synchrotron and SSC Cooling]
{Relativistic Turbulence with Strong Synchrotron and Synchrotron-Self-Compton Cooling}

\author[D.~A.~Uzdensky]{
D.~A.~Uzdensky,$^{1,2}$\thanks{E-mail: uzdensky@colorado.edu}
\\
$^{1}$Center for Integrated Plasma Studies, Physics Department, 
   390 UCB, University of Colorado, Boulder, CO 80309, USA\\
$^{2}$ Institute for Advanced Study, 1 Einstein Dr., Princeton, NJ 08540, USA \\
}

\date{Accepted 2018 March 16. Received 2018 February 26; in original form 2017 March 14}
\pubyear{2018}

\begin{document}
\label{firstpage}
\pagerange{\pageref{firstpage}--\pageref{lastpage}}
\maketitle

% Abstract of the paper
\begin{abstract}
%
%This is a simple template for authors to write new MNRAS papers.
%The abstract should briefly describe the aims, methods, and main results of the paper.
%It should be a single paragraph not more than 250 words (200 words for Letters).
%No references should appear in the abstract.
%
Many relativistic plasma environments in high-energy astrophysics, including pulsar wind nebulae, hot accretion flows onto black holes, relativistic jets in active galactic nuclei and gamma-ray bursts, and giant radio lobes, are naturally turbulent.  The plasma in these environments is often so hot that synchrotron and inverse-Compton (IC) radiative cooling becomes important.  In this paper we investigate the general thermodynamic and radiative properties (and hence the observational appearance) of an optically thin relativistically hot plasma stirred by driven magnetohydrodynamic (MHD) turbulence and cooled by radiation.  We find that if the system reaches a statistical equilibrium where turbulent heating is balanced by radiative cooling, the effective electron temperature tends to attain a universal value $\theta = kT_e/m_e c^2 \sim 1/\sqrt{\tau_T}$, where $\tau_T=n_e\sigma_T L \ll 1$ is the system's Thomson optical depth, essentially independent of the strength of turbulent driving and hence of the magnetic field. This is because both MHD turbulent dissipation and synchrotron cooling are proportional to the magnetic energy density. We also find that synchrotron self-Compton (SSC) cooling and perhaps a few higher-order IC components are automatically comparable to synchrotron in this regime. The overall broadband radiation spectrum then consists of several distinct components (synchrotron, SSC, etc.), well separated in photon energy (by a factor $\sim \tau_T^{-1}$) and roughly equal in power. The number of IC peaks is checked by Klein-Nishina effects and depends logarithmically on $\tau_T$ and the magnetic field. We also examine the limitations due to synchrotron self-absorption, explore applications to Crab PWN and blazar jets, and discuss links to radiative magnetic reconnection. 
\end{abstract}

% Select between one and six entries from the list of approved keywords.
% Don't make up new ones.

%: 
%The full list of MNRAS key words is available here: \\
%% http://oxfordjournals.org/our_journals/mnrasl/for_authors/mnraskey.pdf

%%%%%%%%%%%%%%%%%%%%%%%%%%%%%%%%%%%%%%%%%%%%%%%%%%

%%%%%%%%%%%%%%%%% BODY OF PAPER %%%%%%%%%%%%%%%%

%***********************************************************************************

\section{Introduction}
\label{sec-intro}

Magnetohydrodynamic (MHD) turbulence is a fundamental physical process, ubiquitous in many space and astrophysical plasma environments \citep{Biskamp-book}. In particular, it is widely believed to play an important role in various relativistic high-energy astrophysical plasma flows, such as pulsar wind nebulae (PWN), accretion flows on black holes in X-ray Binaries (XRBs) and active galactic nuclei (AGN), their associated relativistic jets and  giant radio lobes, and Gamma-Ray Burst (GRB) jets and afterglows. These environments are often optically thin and, at the same time, they shine brightly across the electromagnetic spectrum. 
The emission is produced by several radiative processes, most importantly synchrotron and inverse-Compton (IC); the latter in general can be a combination of synchrotron self-Compton (SSC) and external~IC.  The radiative cooling time of the emitting high-energy electrons (and, if present, positrons) is often inferred to be shorter than their travel time from the central engine powering the flow (e.g., an accreting supermassive black hole powering an AGN jet), indicating that substantial energy dissipation and particle acceleration takes place locally.  Since most astrophysical flows are generically magnetized and turbulent, MHD turbulence provides a plausible mechanism for in-situ particle energization powering at least the observed persistent emission --- probably in combination with magnetic reconnection and shocks that may also arise in these flows and that may be responsible for the strongest particle acceleration events powering the brightest high-energy flares. 

The dynamics and statistical properties of non-radiative magnetized plasma turbulence, both incompressible and compressible, including relativistic turbulence \citep[e.g.,][]{Zrake_MacFadyen-2012, Zrake_MacFadyen-2013, Zhdankin_etal-2017}, have been studied theoretically very extensively in the past several decades \citep[e.g.,][]{Goldreich_Sridhar-1995, Biskamp-book, Cho_Lazarian-2003, Cho_etal-2003, Padoan_etal-2004, Zhou_etal-2004, Mueller_Grappin-2005, Boldyrev-2006, Kowal_etal-2007, Howes_etal-2008, Brandenburg_Lazarian-2013, Schekochihin_etal-2009, Zhdankin_etal-2013, Zhdankin_etal-2015a}. 
In contrast, the effects of radiative cooling on MHD turbulence have so far been relatively little explored. A few notable exceptions include several specific astrophysical applications, such as MRI-driven turbulence in accretion disks \cite[e.g.,][]{Hirose_etal-2006, Uzdensky-2013, Jiang_etal-2013}, turbulence in galaxy clusters \citep{Parrish_etal-2010, Kunz_etal-2011}, and turbulence in the interstellar medium (ISM) \citep{Vazquez-Semadeni_etal-1996, Joung_MacLow-2006}.  
In the relativistic case, relevant to systems like PWN, AGN jets and radio-lobes, GRBs, etc., MHD turbulence with strong synchrotron or synchrotron-self-Compton cooling in ultra-relativistically hot plasmas is still very poorly understood.  

These remarks provide a strong motivation for our present study.   
Our goal it to understand the basic properties of relativistic plasmas where turbulent heating is balanced by radiative cooling. In particular, we will focus on the case of continuously driven turbulence (commonly considered in traditional turbulence studies) and will examine the equilibrium thermodynamic and radiative properties of the plasma in the statistical steady state, such as the saturated temperature and the emitted spectrum. 
From a broader perspective, our study aims to advance the emerging field of radiative plasma astrophysics and, we hope, will give a deeper physical insight that will be helpful to understanding high-energy astrophysical systems. 

Our key findings are: 
(1) as long as turbulence is relativistic, $v_{\rm turb} \sim c$, the equilibrium saturated rms particle energy, and hence the effective temperature, is independent of the strength of driving and, correspondingly, of the magnetic field and is just inversely proportional to the square root of the Thomson optical depth across the system; 
(2) SSC and, under certain conditions, a few higher-order IC cooling components are automatically comparable in power to the synchrotron component.

This paper is organized as follows.
We present our analysis in \S~\ref{sec-analysis}. 
In particular, in \S~\ref{subsec-synch} we discuss the hypothetical case of pure synchrotron cooling and derive our main estimate of the equilibrium temperature. 
In \S~\ref{subsec-synch_self-absorption}, we analyze the limitations to our model due to the possibility of synchrotron self-absorption. 
In \S~\ref{subsec-SSC} we consider SSC and other, higher-order IC components. 
In \S~\ref{sec-astro} we discuss astrophysical applications to the Crab nebula and blazar jets, and in \S~\ref{sec-conclusions} we present some additional discussion, outline promising future research directions, and draw our conclusions.

%***********************************************************************************

\section{Analysis}
\label{sec-analysis}

Let us consider driven, statistically steady magnetohydrodynamic turbulence in a plasma with relativistically hot (namely, with the typical rms particle energy greatly exceeding the rest-mass energy, $\gamma_{\rm rms} m_e c^2 \gg m_e c^2$) electrons (in the case of an electron-ion plasma) or electrons and positrons (for pair plasma), subject to strong optically thin radiative cooling.  We will assume that the turbulence is continuously driven at some large scale~$L$, which we tentatively associate with the system size, and will consider the long-term thermal balance established between turbulent heating and radiative cooling.  We will specifically focus on the effects of synchrotron and synchrotron-self-Compton (SSC) cooling, while leaving other radiation mechanisms, such as external IC, Bremsstrahlung, and curvature radiation, to future studies. 
We shall assume that the plasma is optically thin to Thomson scattering, i.e., 
\beq
\tau_T \equiv n \sigma_T L \ll 1 \, ,
\label{eq-tau_T}
\eeq
where $\sigma_T = (8\pi/3)\, r_e^2 \simeq 6.65\times 10^{-25}\,{\rm cm}^2$ is the Thomson cross-section and $r_e \equiv e^2/m_e c^2 \simeq 2.8\times 10^{-13}\,{\rm cm}$ is the classical electron radius. Here, the density $n$ is the total number density of emitting relativistic particles, i.e., the density of electrons in the case of an electron-ion plasma and the combined density of electrons and positrons in the case of a pair plasma. The Thomson optical depth $\tau_T$ is one of the most important dimensionless control parameters in our analysis.

In our description of turbulence we will ignore possible effects due to the intermittency of turbulent dissipation \cite[e.g.,][]{Zhdankin_etal-2013, Zhdankin_etal-2015a, Zhdankin_etal-2015b, Zhdankin_etal-2016a}. We will also ignore the possibility of plasma inhomogeneity, such as may arise due to a synchrotron-cooling instability, which we relegate to a future study, and will just consider the average, gross thermodynamic properties of the fluid in the saturated, statistically steady state. 

Furthermore, in this paper for simplicity we will ignore nonthermal effects, i.e., the effects of a possible very broad spread of particle energies, and will instead think of the particle population as being quasi-thermal, described by a single characteristic particle energy scale. 
It is true that recent first-principles particle-in-cell (PIC) simulations of kinetic turbulence in non-radiative relativistic pair plasmas have unambiguously shown robust nonthermal particle acceleration~\citep{Zhdankin_etal-2017}.  However, the presence of synchrotron and/or IC radiative losses, which for ultra-relativistic particles are proportional to the square of the particle energy and hence which have a stronger effect on more energetic particles, is likely to make the distribution function effectively narrower and thus suppress nonthermal effects to some extent. 
This consideration provides some justification for ignoring the nonthermal effects in our simple model,  although we believe that such effects should be investigated in a more detailed and rigorous future study. 

Since this paper is primarily concerned with the effect of synchrotron and IC radiation losses on ultra-relativistic particles, which are both proportional to the square of the particle energy, it will be convenient for us to characterize the typical energy scale by the root-mean-square (rms) of the energy, expressed in dimensionless form by the rms Lorentz factor, 
\beq
\gamma_{\rm rms} \equiv \sqrt{\overline{\gamma^2}}  \, , 
\label{eq-gamma_rms-def}
\eeq
where $\overline{\gamma^2}$ is the average square of the particle Lorentz factor.
Since we are here interested in ultra-relativistically hot plasmas, we will assume $\gamma_{\rm rms} \gg 1$. 
Although real plasmas under the conditions considered in this paper are likely to be nonthermal, we will sometimes find it convenient to introduce an effective plasma "temperature" $kT = \theta\, m_e c^2$, which we will define in terms of the rms particle energy $\gamma_{\rm rms}\, m_e c^2$ as $kT \equiv (\gamma_{\rm rms}/2\sqrt{3})\, m_e c^2$, where the $1/2\sqrt{3} \simeq 1/3.46$ factor is chosen to mimic the relationship $\gamma_{\rm rms} = 2\sqrt{3} \theta$ found for a thermal (J\"uttner-Maxwell) ultra-relativistically hot plasma. Thus, we shall sometimes use the term "effective temperature" colloquially in qualitative discussions. 

Our primary goal in this study is to find the saturated equilibrium value of $\gamma_{\rm rms}$, established as a result of the balance between turbulent heating and radiative cooling, and then use it to examine radiation signatures in the turbulent statistical steady state. 

%----------------------------------------------------------------------------

\subsection{Turbulent Heating}
\label{subsec-heating}

As is normally expected in MHD turbulence, we will assume that the amount of turbulent magnetic and bulk kinetic energy dissipated per unit volume in one large-scale turbulent eddy turnover time $\tau_{\rm turb}=L/v_{\rm turb}$ (corresponding to a characteristic turbulent velocity $v_{\rm turb}$ at the energy-containing scale~$L$) constitutes a substantial fraction of the turbulent magnetic energy density~$B^2/8\pi$.  
We shall denote this fraction by a constant dimensionless parameter~$\kappa$, which we will assume to be of order unity. Then, the turbulent energy dissipation rate per unit volume can be written as 
\beq
\mathcal{E}_{\rm heat} = \kappa\, {B^2\over 8\pi} \, \tau_{\rm turb}^{-1} =  
\kappa\, {B^2\over 8\pi} \, {v_{\rm turb}\over L} = \kappa\, {B^2\over 8\pi} \, \beta_{\rm turb} {c\over L} \, , 
\label{eq-turb-heat}
\eeq
where $B$ is the rms turbulent magnetic field and $\beta_{\rm turb} \equiv v_{\rm turb}/c$.

Most of the analysis presented in this paper is general, in the sense that it does not require the turbulent motions to be relativistic and thus does not make any strong assumptions about~$\beta_{\rm turb}$.  However, the specific case of {\it relativistic turbulence}, in which $v_{\rm turb} \sim c$,  $\beta_{\rm turb}  \sim 1$, and $\tau_{\rm turb}$ is comparable to the light crossing time~$L/c$, is of particular interest and has strong astrophysical motivation. This case is realized when the supplied driving power is sufficiently large; the corresponding validity condition of this regime can be expressed as follows.
%In this paper we will be primarily interested in the case of relativistic turbulence, where $v_{\rm turb} \sim c$ and so $\beta_{\rm turb}  \sim 1$; then, $\tau_{\rm turb}$ becomes comparable to the light crossing time~$L/c$.  
%This is true when the supplied driving power is sufficiently large and the validity condition of this assumption can be expressed as follows. 
MHD turbulence is Alfv\'enic, characterized by energy equipartition between turbulent kinetic and magnetic energy densities. Therefore, $v_{\rm turb} \sim V_A$, where~$V_A$ is the Alfv\'en speed corresponding to the turbulent magnetic field~$B$ (we shall assume that there is no strong background guide magnetic field~$B_g$). 
For a relativistically hot plasma $V_A$ is given by $V_A = c\, \sigma^{1/2}/(1+ \sigma)^{1/2}$, 
where $\sigma \equiv B^2/4\pi h$ is the plasma magnetization expressed in terms of the relativistic enthalpy density 
$h \equiv (4/3) U_{\rm pl} \equiv (4/3)\, n \bar{\gamma} m_e c^2 $.
Thus, the condition $v_{\rm turb} \sim V_A \sim c$ is equivalent to $\sigma \gtrsim 1$; i.e., the turbulent magnetic field  and hence turbulent driving need to be sufficiently strong, so that the magnetic energy density is at least comparable to  the plasma internal energy density $U_{\rm pl}= n \bar{\gamma} m_e c^2$, 
\beq
{B^2\over{8\pi}}  \gtrsim  n \bar{\gamma} m_e c^2  \, .
\eeq
This corresponds to a lower limit on the supplied power per unit volume (and therefore on the volumetric radiative luminosity of the system):
\beq
\mathcal{E}_{\rm heat} \gtrsim \kappa\, {v_{\rm turb}\over L} \, U_{\rm pl} \, .
\eeq
This means that the system needs to be in a strong-cooling regime, so that the cooling time is shorter than~$\tau_{\rm turb}=L/v_{\rm turb}$, or, equivalently, the amount of stored internal energy is less than the amount supplied per eddy turnover time.

%----------------------------------------------------------------------------

\subsection{Synchrotron cooling}
\label{subsec-synch}

If a statistical steady state is achieved, the plasma heating rate (\ref{eq-turb-heat}) is balanced, on average, by radiative cooling.  Let us first consider, as an exercise, the case of pure synchrotron cooling while postponing the discussion of~SSC until~\S~\ref{subsec-SSC}. As mentioned above, for simplicity we shall assume that there is no strong background magnetic field, so that the synchrotron emission is essentially entirely due to the turbulent magnetic field~$B$.
We shall also assume that the distribution of particle pitch angles $\alpha$ relative to the local direction of the magnetic field is isotropic, so that $\overline{\sin^2\alpha}=2/3$. Then, the synchrotron cooling rate per unit volume for a relativistically hot plasma is 
\beq
\mathcal{E}_{\rm cool} \simeq \mathcal{E}_{\rm synch} \simeq 
2\, n \sigma_T c\, {B^2\over 8\pi} \, \overline{\gamma^2} \ \overline{\sin^2\alpha} \simeq
{4\over 3} \, n \sigma_T c\, {B^2\over 8\pi} \,\gamma_{\rm rms}^2 \, , 
\label{eq-cool-synch-1}
\eeq
where we have neglected any possible correlation between the local values of the magnetic field and~$\gamma_{\rm rms}$. 
We can rewrite this expression in terms of $\tau_T=n\sigma_T L$ as
\beq
 \mathcal{E}_{\rm synch} \simeq \tau_T \, {c\over L}\, {B^2\over 6\pi}\, \gamma_{\rm rms}^2 \, ,
\label{eq-cool-synch-2}
\eeq
which can also be rewritten in terms of the effective plasma temperature~$\theta = \gamma_{\rm rms}/2\sqrt{3}$ as
$ \mathcal{E}_{\rm synch} \simeq (2/\pi)\, (c/ L)\, \tau_T\, B^2\, \theta^2$.

Importantly, both the dissipation rate (\ref{eq-turb-heat}) and the synchrotron cooling rate (\ref{eq-cool-synch-2}) are explicitly proportional to the turbulent magnetic energy density~$B^2/8\pi$. Thus, when we equate the two rates to establish the thermal balance, 
\beq
\mathcal{E}_{\rm heat} =  \mathcal{E}_{\rm synch} \, , 
\label{eq-heat_cool-balance}
\eeq
the magnetic field just cancels out, as does the light crossing time~$L/c$, and we get a very simple expression for the equilibrium plasma temperature [ignoring, in light of crudeness of our model, the numerical factor $(3/4)^{1/2} \simeq 0.87$]: 
\beq
2\sqrt{3}\,\theta = \gamma_{\rm rms} \simeq  \sqrt{\kappa \beta_{\rm turb}\over{\tau_T}}  \, .
\label{eq-theta-tau_T}
\eeq 
This is a simple, but remarkable result. It means that the effective plasma temperature in a steady state is independent of the strength of turbulent driving and hence of the resulting turbulent magnetic field. This is because, in the case of synchrotron cooling, the magnetic field plays a dual role: it is the main agent for both heating and radiative cooling. Thus, if one increases the turbulent driving rate, one makes the saturated magnetic field stronger, but this also makes the cooling strength higher in proportion, so that the resulting saturated equilibrium temperature is unchanged. 
In particular, in the case of relativistic turbulence, $\kappa \beta_{\rm turb} \sim 1$, as we see from the above simple expression, this equilibrium effective temperature $\theta$ only depends on the Thomson optical depth across the system (or, more accurately, across the turbulent driving scale). We also see that our initial assumption of optically thin plasma, $\tau_T \ll 1$, is consistent with our other assumption that the plasma is ultra-relativistically hot, $\gamma_{\rm rms} \gg 1$. 

%----------------------------------------------------------------------------------------------------

Next, using the above estimate (\ref{eq-theta-tau_T}) for the temperature, we can estimate several other important plasma parameters and scales. For example, we can write the electron collisionless skin depth (which for relativistically hot plasma coincides with the Debye length~$\lambda_D$) as
\beq
d_e \equiv {c\over{\omega_{pe}}} = \sqrt{{\bar{\gamma} m_e c^2}\over{4\pi n e^2}} = 
\sqrt{{\bar{\gamma}}\over{4\pi n r_e}} \sim \sqrt{r_e L} \, (\kappa\beta_{\rm turb})^{1/4}\, \tau_T^{-3/4} \, .
\eeq
The basic scaling $d_e \sim (r_e L)^{1/2}$ at a finite optical depth is similar to what one finds in coronae of accreting black holes \cite{Goodman_Uzdensky-2008}.
The corresponding plasma parameter, 
$N_D \equiv n \lambda_D^3 \simeq n d_e^3 \sim (n r_e^3)^{-1/2}\, (\kappa \beta_{\rm turb})^{3/4}\, \tau_T^{-3/4}$ is automatically very large, and so the plasma approximation is valid.  

%\medskip
Next, we can estimate the characteristic critical synchrotron frequency for an rms particle 
\beq
\omega_{\rm synch} \equiv 2\pi \nu_{\rm synch} = 
{3\over 2}\, \Omega_c \overline{\gamma^2}\ \overline{\sin\alpha} \simeq 
{3\pi\over 8}\, \Omega_c \overline{\gamma^2}  \sim \Omega_c \kappa \beta_{\rm turb} \tau_T^{-1} \, , 
\label{eq-nu-synch}
\eeq
where $\Omega_c \equiv eB/m_e c$ is the classical non-relativistic electron cyclotron frequency, and where we again assumed an isotropic pitch-angle distribution, $\overline{\sin\alpha} = \pi/4$, and used the estimate~(\ref{eq-theta-tau_T}).  
The corresponding synchrotron wavelength can be expressed as
\beq
\lambda_{\rm sync} = {c\over\nu_{\rm sync}} = {16\over 3}\, {{\rho_0} \over{\gamma_{\rm rms}^2}} 
\sim {16\over 3}\, {\rho_0}\, \tau_T  (\kappa \beta_{\rm turb})^{-1}  \, , 
\label{eq-lambda-synch}
\eeq
where $\rho_0 \equiv c/\Omega_c = m_e c^2/eB$. 

Next, when we describe the characteristic synchrotron photon energy,  
\beq
\epsilon_{\rm synch} = \hbar \omega_{\rm synch} = 
{3\over 2}\, \hbar\, \Omega_c \overline{\gamma^2} \sim 
\hbar\, \Omega_c \kappa \beta_{\rm turb} \tau_T^{-1} \, , 
\label{eq-epsilon-synch}
\eeq
we find that it is often convenient to express the magnetic field in a dimensionless form by normalizing it to the critical quantum (Schwinger) field
\beq
B_Q \equiv {{m_e^2 c^3}\over{e \hbar}} \simeq 4.4 \times 10^{13} \, {\rm G}, 
\eeq
[corresponding to $\hbar \Omega_c = m_e c^2$], and by defining  
\beq
b \equiv {B\over{B_Q}} \, . 
\eeq
With this definition we can write 
\beq
\hbar \Omega_c = b m_e c^2 \, ,
\label{eq-cyclotron-energy}
\eeq 
and so 
\beq
\epsilon_{\rm synch} \sim m_e c^2\,  \kappa \beta_{\rm turb}\, {b\over{\tau_T}} \, .
\eeq

%***********************************************************************************

\subsection{Synchrotron Self-Absorption}
\label{subsec-synch_self-absorption}

When the optical depth of the emitting plasma is not sufficiently small, the emitted synchrotron radiation may suffer significant self-absorption before it can escape the system, thereby invalidating our assumption of optically thin synchrotron cooling. In particular, synchrotron self-absorption (SSA) becomes non-negligible when the spectral density of the synchrotron radiation at the relevant frequency, namely, the peak frequency of the synchrotron radiation $\nu_{\rm synch,peak} \simeq 0.29 \nu_{\rm synch}$, becomes comparable to the blackbody intensity corresponding to the plasma temperature $kT = \theta m_e c^2$~\citep{Longair-2011}.
%(e.g., M.~Longair, "High-Energy Astrophysics", p.~217). 
If this happens, the rate of synchrotron radiative losses is reduced, necessitating a modification to our model. 
In this subsection we will derive the conditions under which the SSA effects can be neglected. 
Although the general spirit of this paper is to ignore numerical factors of order unity, in this subsection we shall attempt to retain them as much as possible because we find that they can quickly escalate to rather large values. 

Let us for simplicity assume that the plasma is a sphere of radius~$L$, shining isotropically with a uniform volumetric synchrotron emissivity $j_{\rm sync}(\nu)$, and thus estimate the synchrotron intensity at a given frequency $\nu$ as
\beq
I_{\nu,\rm sync}(\nu)  \sim {{4 \pi L^3/3}\over{4\pi L^2}}\, j_{\rm sync}(\nu) \simeq  {L\over 3} \, j_{\rm sync}(\nu) \, . 
\label{eq-synch-intensity}
\eeq
Assuming isotropic distribution of particle pitch angles, corresponding to $\overline{\sin\alpha} = \pi/4$, we can evaluate the synchrotron emissivity $j_{\rm sync}(\nu) \simeq (\sqrt{3}/2\pi)\, n r_0 e B \, \overline{\sin\alpha} \, F(\nu/\nu_{\rm synch})$
as 
\beq
j_{\rm sync}(\nu) \simeq {\sqrt{3}\over{8}}\, n r_0 e B\, F(\nu/\nu_{\rm synch}) \, , 
\eeq
where $F(x)$ is the standard synchrotron spectrum function \cite[e.g.,][]{Ginzburg_Syrovatskii-1965}. 
This emissivity can then be conveniently expressed in terms of the total synchrotron power density $\mathcal{E}_{\rm synch}$, given by~(\ref{eq-cool-synch-1}), and the critical synchrotron frequency  $\nu_{\rm synch}$, given by (\ref{eq-nu-synch}), as 
\beq
j_{\rm sync}(\nu) \simeq {81\sqrt{3}\over{256 \pi}}\, {\mathcal{E}_{\rm synch}\over{\nu_{\rm synch}}} \, F(\nu/\nu_{\rm synch}) \simeq  
0.17 \, {\mathcal{E}_{\rm synch}\over{\nu_{\rm synch}}} \, F(\nu/\nu_{\rm synch}) \, .
\eeq

Substituting this result into our estimate (\ref{eq-synch-intensity}) for the synchrotron intensity $I_{\nu,\rm sync}(\nu)$ and evaluating it at $\nu_{\rm synch,peak} \simeq 0.29\,\nu_{\rm synch}$, where $F(0.29)\approx 1$ \cite[e.g.,][]{Ginzburg_Syrovatskii-1965}, 
we get
\beq
I_{\nu,\rm sync}(\nu_{\rm synch,peak}) \simeq {27\sqrt{3}\over{256 \pi}}\, L\, {\mathcal{E}_{\rm synch}\over{\nu_{\rm synch}}} \simeq 
0.06\, L\, {\mathcal{E}_{\rm synch}\over{\nu_{\rm synch}}} \, .
\eeq

Now, as long as synchrotron cooling is balanced by turbulent heating, so that equation~(\ref{eq-heat_cool-balance}) holds, we can replace $\mathcal{E}_{\rm synch}$ with $\mathcal{E}_{\rm heat}$ and, using equation~(\ref{eq-turb-heat}), obtain
\beq
I_{\nu,\rm sync}(\nu_{\rm synch,peak}) \simeq {27\sqrt{3}\over{256 \pi}}\, L\, {\mathcal{E}_{\rm heat}\over{\nu_{\rm synch}}} \simeq 
{27\sqrt{3}\over{256 \pi}}\, {B^2\over 8\pi} \, \kappa\beta_{\rm turb} \, \lambda_{\rm synch} 
\simeq 0.06\, {B^2\over 8\pi} \, \kappa\beta_{\rm turb} \, \lambda_{\rm synch} \, , 
\label{eq-I_synch}
\eeq
where $\lambda_{\rm synch} \equiv c/\nu_{\rm synch}$ is the critical synchrotron wavelength given by~(\ref{eq-lambda-synch}). 

On the other hand, the blackbody radiation intensity at the same frequency $\nu = \nu_{\rm synch,peak}$, assuming that $h\nu_{\rm synch,peak} \ll kT$ so that the Rayleigh-Jeans regime applies, is
\beq
I_{\nu,\rm bb}(\nu_{\rm synch,peak}) = {{2 kT \nu_{\rm synch,peak}^2}\over{c^2}} =
{2 kT \over{\lambda_{\rm synch,peak}^2}} \simeq 
0.29^2 \, {2 kT \over{\lambda_{\rm synch}^2}}  \simeq 0.17\, {kT \over{\lambda_{\rm synch}^2}}\, . 
\label{eq-I_bb}
\eeq

Thus, by comparing equations~(\ref{eq-I_synch}) and~(\ref{eq-I_bb}), we get a very simple condition for synchrotron self-absorption to be negligible: 
\beq
I_{\nu,\rm bb}(\nu_{\rm synch,peak}) \gg I_{\nu,\rm sync}(\nu_{\rm synch,peak})  \quad \Rightarrow \quad 
kT \gg  {27\sqrt{3}\over{512 \pi}}\, {1\over{0.29^2}}\, {B^2\over{8\pi}} \, \lambda_{\rm synch}^3\, \kappa \beta_{\rm turb} \simeq {B^2\over{8\pi}} \, \lambda_{\rm synch}^3\, {\kappa \beta_{\rm turb}\over{3}}  \, .
\label{eq-noSSA-condition-1}
\eeq
Physically, this roughly means that the temperature is greater than the magnetic energy contained in one cubic synchrotron wavelength. This condition can be recast in a compact form in terms of plasma beta,  $\beta \equiv 8\pi n k T /B^2$, and the number of plasma particles per cubic critical synchrotron wavelength
\beq
\beta \gg {{\kappa \beta_{\rm turb}} \over{3}}\, n\lambda_{\rm synch}^3  \, .
\eeq

Next, using equation~(\ref{eq-lambda-synch}) and expressing 
$\rho_0^3 B^2/8\pi = m_e c^2\, e/(8\pi B r_e^2) = m_e c^2/(8\pi \alpha b)$, 
where we have introduced the fine structure constant $\alpha \equiv e^2/\hbar c \simeq 1/137$, we can write 
\begin{eqnarray}
\lambda_{\rm sync}^3 \, {B^2\over{8\pi}} = 
\biggl({16\over 3}\biggr)^3 \, \gamma_{\rm rms}^{-6}\, \rho_0^3 \, {B^2\over{8\pi}} = 
%\biggl({16\over 3}\biggr)^3 \, \gamma_{\rm rms}^{-6}\, m_e c^2\, {e\over{8\pi B r_e^2}} =
\biggl({16\over 3}\biggr)^3 \, \gamma_{\rm rms}^{-6}\, {{m_e c^2}\over{8\pi \alpha b}} = 
{512\over{27\pi \alpha}}\,  \gamma_{\rm rms}^{-6}\, b^{-1}\, m_e c^2 \simeq 
{6\over \alpha}\, \gamma_{\rm rms}^{-6}\, b^{-1}\, m_e c^2 \simeq
830\, \gamma_{\rm rms}^{-6}\, b^{-1}\, m_e c^2 \, .
\end{eqnarray}
Substituting this expression into the condition~(\ref{eq-noSSA-condition-1}), our condition that SSA can be neglected becomes
\beq
\theta = {kT\over{m_e c^2}} \gg  
{\sqrt{3}\over{0.29^2 \pi^2}}\, {1\over \alpha}\, \gamma_{\rm rms}^{-6}\, b^{-1} \kappa \beta_{\rm turb} \simeq 
{2\over \alpha}\, \gamma_{\rm rms}^{-6}\, b^{-1} \kappa \beta_{\rm turb}  \, .
\label{eq-noSSA-condition-2}
\eeq

Finally, recalling our definition of the effective temperature, $k T = \theta m_e c^2 = \gamma_{\rm rms} m_e c^2/(2\sqrt{3})$, our condition that SSA can be neglected becomes 
\beq
b \, \gamma_{\rm rms}^{7} \gg {6\over{0.29^2 \pi^2}}\, \alpha^{-1}\, \kappa\beta_{\rm turb} \simeq 
{7.2\over{\alpha}}\, \kappa\beta_{\rm turb} \simeq 1000\, \kappa\beta_{\rm turb} \, , 
\label{eq-noSSA-condition-3}
\eeq
and, using the estimate (\ref{eq-theta-tau_T}), the condition becomes 
\beq
b \gg 1000\, (\kappa\beta_{\rm turb})^{-5/2} \, \tau_T^{7/2} \, , \quad {\rm or} \quad
\tau_T \ll 0.14 \, b^{2/7} \, (\kappa\beta_{\rm turb})^{5/7}\, .
\label{eq-noSSA-condition-4}
\eeq

Thus, for example, for the conditions of the Crab Nebula, with $B \simeq 300 \mu{\rm G}$ and hence $b \sim 10^{-17}$, this requirement becomes (taking $\kappa\beta_{\rm turb} \sim 1$) $\tau_T \ll 10^{-4}$, or, equivalently, $n \ll 50\, {\rm cm^{-3}} \, L_{\rm pc}$, where $L_{\rm pc}$ is the size of the Nebula in parsec. 
This condition is easily satisfied since one expects $\tau_T \sim 10^{-13}$ (see \S~\ref{sec-astro}).

%***********************************************************************************

\subsection{Synchrotron-Self-Compton Cooling}
\label{subsec-SSC}

\begin{figure}
\begin{center}
\includegraphics[width=12.cm]{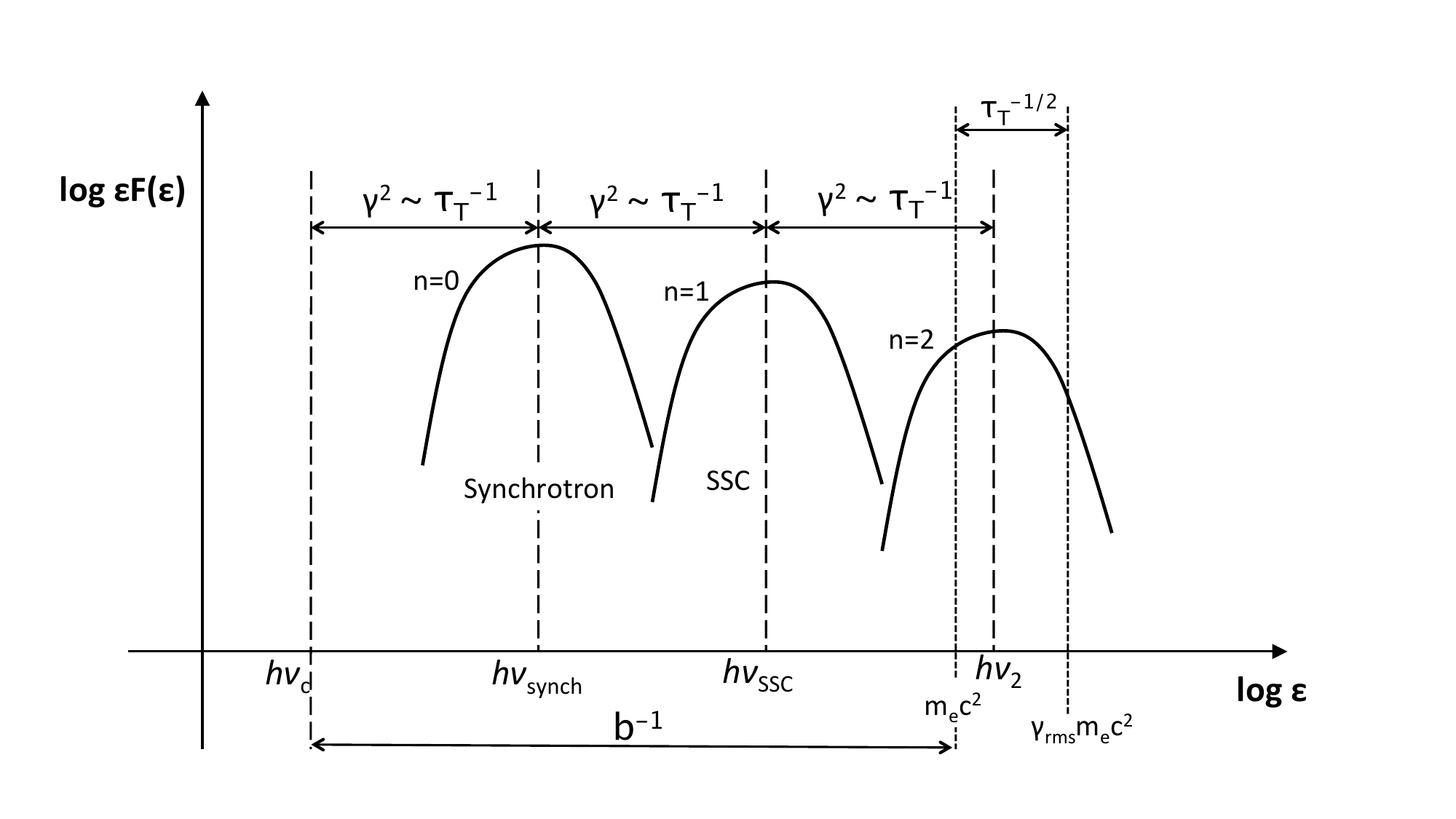}
\end{center}
\caption{Qualitative schematic sketch of the broad-band spectrum of a relativistically turbulent plasma with strong radiative cooling, showing synchrotron, SSC, and one more higher-order IC emission components ($N=2$).
\label{fig-spectrum}
}
\end{figure}

Synchrotron cooling is not the only radiative cooling mechanism operating in relativistic astrophysical plasmas. 
One should question the validity of the pure-synchrotron assumption and investigate the possible role of synchrotron-self-Compton (SSC) cooling, as well as its "descendants", i.e., higher-order IC components. We will carry out such an analysis in this subsection, while leaving an investigation of external IC cooling to a future study.
We will show here that, when a relativistically-turbulent ($\kappa\beta_{\rm turb} \sim1$) optically-thin ($\tau_T \ll 1$) system reaches a saturated steady state, where turbulent heating is balanced by the total radiative cooling, it then follows automatically that the SSC power $\mathcal{E}_{\rm SSC}$ is comparable to the synchrotron power~$\mathcal{E}_{\rm synch}$. But this is not all! Since $\mathcal{E}_{\rm SSC} \sim \mathcal{E}_{\rm synch}$, the same relationship holds between their corresponding radiation energy densities, and therefore the resulting SSC photons are basically just as efficient as the synchrotron ones in cooling the plasma through IC scattering.  This sets up the conditions somewhat similar to (but not exactly the same as) what is known as the inverse Compton catastrophe \citep{Kellermann_Pauliny-Toth-1969}.
Thus, one can go the next level and find that the IC scattering of SSC photons again yields a cooling contribution that is automatically comparable to $\mathcal{E}_{\rm SSC}$ and~$\mathcal{E}_{\rm synch}$.  And one can repeat such an estimate again and again, in discrete steps, or stages, each time finding a separate IC radiation component with nearly the same power, and a peak photon energy by a factor of $\gamma_{\rm rms}^2 \sim \tau_T^{-1}$ higher, than those of the previous component.
The overall broadband spectrum of such a system thus involves, in addition to the synchrotron component, several discrete IC peaks, spaced logarithmically equidistantly in photon energy and with gradually decreasing cooling power.  The first IC peak is the SSC itself, the second is due to the IC scattering of the SSC photons, etc.. This sequence continues until the IC photon energy becomes comparable to the typical electron energy $\gamma_{\rm rms} m_e c^2$ (see Fig.~\ref{fig-spectrum}).

%----------------------------------------------------------------------

%----------------------------------------------------------------------

As a first step, let us modify the analysis in \S~\ref{subsec-synch} slightly by introducing a parameterization for the anticipated reduction of the synchrotron contribution to the overall radiative cooling~$\mathcal{E}_{\rm rad,tot}$.  Namely, denoting the synchrotron cooling fraction by 
\beq
q_0 \equiv {\mathcal{E}_{\rm synch}\over{\mathcal{E}_{\rm rad,tot}}} \leq 1 \, , 
\eeq
which we shall regard as a dimensionless constant of order unity, and repeating the steps outlined in \S~\ref{subsec-synch}, we can write (again for simplicity ignoring the factor of 3/4) 
\beq
\mathcal{E}_{\rm synch} \simeq n c \sigma_T \, {B^2\over{8\pi}} \, \overline{\gamma^2} \simeq
q_0 \, \mathcal{E}_{\rm rad,tot} = q_0 \, \mathcal{E}_{\rm heat} =
  q_0 \kappa\, {B^2\over{8\pi}} \, {\beta_{\rm turb} c\over{L}} \, . 
\eeq
From this we can again estimate the rms-average particle Lorentz factor, 
$\gamma_{\rm rms} \equiv (\overline{\gamma^2})^{1/2}$, in terms of $\tau_T$ as
\beq
\tau_T \overline{\gamma^2} \simeq q_0 \kappa \beta_{\rm turb}  = O(1) \, .
\label{eq-theta-tau_SSC}
\eeq
Thus, we anticipate that the main result of our analysis in \S~\ref{subsec-synch} --- the scaling $\gamma_{\rm rms} \sim \tau_T^{-1/2}$ --- will still hold even in the presence of SSC and higher-order IC cooling.

Now let us discuss the overall structure of the broadband radiation spectrum. 
In a relativistically hot plasma, the IC radiation components are well separated in photon energy, because the characteristic photon energy of the $n$th IC component, $\epsilon_n$, is by a factor $\overline{\gamma^2}\gg 1$ higher than that of the $(n-1)$th-generation photons, which act as the soft seed photons for the $n$th component, i.e, 
\beq
\epsilon_n \simeq \overline{\gamma^2}\, \epsilon_{n-1} \, .
\eeq
Tracing this sequence recursively all the way down to $n=0$ (synchrotron), we obtain
\beq
\epsilon_n \simeq (\overline{\gamma^2})^{n}\, \epsilon_0 =
(\overline{\gamma^2})^{n}\, \epsilon_{\rm synch} \simeq 
(\overline{\gamma^2})^{n+1}\, \hbar\Omega_c  \sim \tau_T^{-(n+1)} \, \hbar\Omega_c \, ,
\eeq
where we used the estimate~(\ref{eq-theta-tau_SSC}) and ignored factors of order unity such as $q_0 \kappa\beta_{\rm rec} = O(1)$. For example, $n=1$ corresponds to the SSC component, 
$\epsilon_{\rm SSC} \sim \tau_T^{-1} \, \epsilon_{\rm synch} \sim \tau_T^{-2} \, \hbar\Omega_c$.

In reality, this progression of IC components does not continue indefinitely. 
It effectively terminates when the corresponding IC photon energy becomes comparable to the typical electron energy or, equivalently, when the IC process enters the Klein-Nishina regime. 
This happens at $n=N$ such that 
\beq 
\epsilon_N \simeq (\overline{\gamma^2})^{N+1}\, \hbar\Omega_c \sim m_e c^2 \gamma_{\rm rms} \, , 
\eeq
i.e., when 
\beq
 (\overline{\gamma^2})^{N+1/2} \sim {m_e c^2 \over{\hbar\Omega_c}} = b^{-1}\, ,
\eeq
where we have used Eq.~(\ref{eq-cyclotron-energy}) to express $\hbar \Omega_c$ in terms of~$b = B/B_Q$.

This yields a simple estimate for the number of expected IC components: 
\beq
N \simeq -\, {\ln b\over{\ln(\overline{\gamma^2})}} - {1\over 2} \, , 
\eeq
and, recalling our expression (\ref{eq-theta-tau_SSC}) for $\overline{\gamma^2}$, and neglecting $|\ln (q_0 \kappa\beta_{\rm turb})|$ compared to $|\ln \tau_T|$, we arrive at 
\beq
N \sim {\ln b\over{\ln \tau_T}} - {1\over 2}  = {|\ln b|\over{|\ln \tau_T}|} - {1\over 2}  \, .
\label{eq-N}
\eeq
Thus we see that the number of IC radiation components depends only logarithmically on $b$ and~$\tau_T$. 
For example, we find that the presence of at least the SSC component ($N\geq 1$) requires, crudely, $|\ln b| \geq 1.5| \ln \tau_T|$ or $b \leq \tau_T^{3/2}$. Many IC components ($N \gg 1$) are expected only when $|\ln b| \gg |\ln \tau_T|$, i.e., for a very weak magnetic field and only modestly small optical depth. 

If one considers systems with ever increasing (but still small) optical depth at fixed~$b$, the equilibrium electron temperature decreases and becomes comparable to $m_e c^2$  as $\tau_T$ approaches~1. Then, the number of IC components increases, the separation between them becomes smaller, and they eventually blur together into one smooth spectrum at a finite optical depth. The picture then becomes similar to what happens in Componizing coronae of accreting black holes \citep{Sunyaev_Titarchuk-1980}.

On the other hand, however, as one raises the optical depth and hence lowers the plasma temperature, one may become concerned about synchrotron self-absorption, which we considered in~\S~\ref{subsec-synch_self-absorption}.
By combining the estimate~(\ref{eq-N}) with the condition~(\ref{eq-noSSA-condition-4}) that SSA can be neglected, $b \gg 1000\, \tau_T^{7/2}$ and so $\ln b > 3.5 \ln \tau_T + 7$, we find 
(taking into account that $\ln \tau_T < 0$), 
\beq
N < {{\ln \tau_T^{7/2} +7}\over{\ln\tau_T}} - {1\over 2} = 3 + {7\over{\ln\tau_T}} \simeq 3 + {3\over{\log_{10} \tau_T}} = 
3 - {3\over{|\log_{10} \tau_T}|}\, , 
\label{eq-N-SSA}
\eeq
i.e, if SSA can be neglected, the system's broad-band spectrum should have no more than a couple IC components in addition to SSC. 

%\medskip
We now turn to estimating the relative contributions of the different IC components to the total cooling rate, and hence 
to the observable broad-band radiation spectrum. 
In order to flesh out the picture more quantitatively, let us write the heating-cooling balance as
\beq
\mathcal{E}_{\rm heat} = \mathcal{E}_{\rm cool} = \mathcal{E}_{\rm rad,tot} = 
\sum \limits_{n=0}^N \mathcal{E}_{\rm rad}^{(n)},  
\eeq
where, once again, $n=0$ corresponds to synchrotron radiation, $n=1$ to SSC, and $N$ is the total number of IC components.  We will parametrize the relative importance of the different radiation components by the dimensionless coefficients 
\beq
q_n \equiv {{\mathcal{E}_{\rm rad}^{(n)}}\over{\mathcal{E}_{\rm rad,tot}}} < 1, \quad n = 0, ..., N \, , 
\eeq
which satisfy the normalization condition 
\beq
\sum \limits_{n=0}^N  q_n = 1 \, .
\label{eq-Sum_q_n-1}
\eeq

Our goal here is to estimate the coefficients~$q_n$ and we will accomplish this by constructing a recursive relationship between them. First, we can write the volumetric power of each of the successive IC components in terms of the radiation energy density $U_{\rm rad}^{(n)}$ of the preceding component as
\beq
\mathcal{E}_{\rm rad}^{(n+1)} \sim 
 n c \sigma_T U_{\rm rad}^{(n)}\, \overline{\gamma^2} \quad n = 0, ..., N-1 \, ,
\eeq
where for simplicity we dropped the numerical pre-factor of~4/3.
Then, estimating (while again ignoring geometrical factors of order unity) 
\beq
U_{\rm rad}^{(n)} \sim \mathcal{E}_{\rm rad}^{(n)}\, {L\over{c}} \, , 
\eeq
expressing the result in terms of $q_n$-s, and using equation~(\ref{eq-theta-tau_SSC}), we obtain 
\beq
q_{n+1} \sim \tau_T \overline{\gamma^2} q_n \simeq (\kappa q_0 \beta_{\rm turb})\, q_n  \, .
\eeq
Thus, the sequence of the coefficients $q_n$ forms a geometric progression, 
\beq
q_n \simeq r^n \, q_0 \, , 
\label{eq-geom-progression}
\eeq
with the common ratio 
\beq
r \simeq \tau_T \overline{\gamma^2} \simeq q_0 \kappa \beta_{\rm turb} < 1 \, . 
\eeq

Thus, since we formally regard the combination $r = q_0 \kappa \beta_{\rm turb}$ as a finite number of order unity, in the sense that it does not scale with either $b$ or~$\tau_T$, we see that the IC contributions $\mathcal{E}_{\rm rad}^{(n)}$ to the total cooling rate are formally comparable to each other and to the synchrotron contribution. This in turn means that, on the one hand, the SSC and higher-order IC components are not negligible but, on the other hand, do not dominate over synchrotron emission. Therefore, they may change the estimates obtained in the previous sections only by numerical factors of order unity, implying that all our conclusions based on the pure-synchrotron model remain qualitatively valid. 
A similar statement can also be made about the corresponding radiation energy densities, as they all turn out to be comparable to the magnetic energy density $U_{\rm magn} = B^2/8\pi$, i.e., 
\beq
U_{\rm rad}^{(n)}  \sim r U_{\rm rad}^{(n-1)} \sim ... \sim r^{n-1} U_{\rm SSC} \sim r^{n} U_{\rm synch} \sim r^{n+1} U_{\rm magn} \, .
\eeq

While equation~(\ref{eq-geom-progression}) expresses the IC coefficients $q_n$ in terms of the synchrotron coefficient~$q_0$, the latter is found by using the normalization condition~(\ref{eq-Sum_q_n-1}): 
\beq
\sum \limits_{n=0}^N  q_n   = q_0\, {{1-r^{N+1}}\over{1-r}}  = 1 \, .
\label{eq-Sum_q_n-2}
\eeq

Let us now consider a couple of illustrative limiting cases. 
First, in the limit of non-relativistic turbulence, $\kappa \beta_{\rm turb} \ll 1$ and hence $r \ll 1$, we see from equation~(\ref{eq-Sum_q_n-2}) that $q_0 \simeq 1$ for any $N$ and so synchrotron cooling dominates over the SSC and the other IC components, whose power then declines rapidly with~$n$, see equation~(\ref{eq-geom-progression}). 

Second, for arbitrary $\kappa \beta_{\rm turb}$, in the case of many active IC components, $N \gg 1$ (although in this case one may have to take into account synchrotron self-absorption as we have shown above, see equation~\ref{eq-N-SSA}), we can neglect $r^{N+1}$ in equation~(\ref{eq-Sum_q_n-2}) and get
\beq
q_0 \simeq 1-r = 1 - q_0 \kappa \beta_{\rm turb} \quad \Rightarrow \quad 
q_0  = {1\over{1+ \kappa \beta_{\rm turb}}} \, .
\label{eq-q_0-large-N}
\eeq
This leads to a situation analogous to the inverse Compton cooling catastrophe, explored previously in the context of extragalactic black-hole-powered radio-sources \cite[see, e.g.,][]{Kellermann_Pauliny-Toth-1969}.
As one can see from equation~(\ref{eq-q_0-large-N}), in the limit of strongly-relativistic turbulence, $\kappa \beta_{\rm turb} \simeq 1$, we obtain $q_0 \simeq r \simeq 1/2$, and so synchrotron radiation is responsible for roughly a half of the overall radiative losses, while the power of each successive IC component drops by a factor  of about~2 relative to the previous component. And in the non-relativistic limit, $\kappa \beta_{\rm turb} \ll 1$, $q_0 \simeq 1$ and then $r \simeq \kappa \beta_{\rm turb} \ll 1$, i.e., synchrotron cooling dominates, in agreement with our general (arbitrary-$N$) conclusion above. 

Third, let us consider the case where only the SSC is present, $N=1$. 
Then, from Eq.~(\ref{eq-Sum_q_n-2}),
\beq
q_0 \, (1-r^2) = 1-r  \quad \Rightarrow \quad  q_0\, (1+r) = 1 \, , 
\eeq
with the only positive solution given by
\beq
q_0 = {{\sqrt{1+4 \kappa \beta_{\rm turb}} - 1} \over{2 \kappa \beta_{\rm turb}}} \, .
\eeq
In the non-relativistic limit $\kappa \beta_{\rm turb} \ll 1$ we once again get $q_0 \simeq 1$ and 
$r = \kappa \beta_{\rm turb} \ll 1$; and in the strongly-relativistic limit $\kappa \beta_{\rm turb} \simeq 1$, 
we see that $q_0$ is related to the golden mean, namely, $q_0 = (\sqrt{5} - 1)/2 \simeq 0.62 $.

%----------------------------------------------------------------------------------------------------

%***********************************************************************************

\section{Astrophysical Implications}
\label{sec-astro}

Brightly shining, relativistically hot, optically thin plasmas are encountered throughout the universe and they are generically expected to be turbulent. Examples of such environments include PWN, hot accretion flows onto black holes, relativistic jets powered by super-massive black holes in AGN including blazars, giant radio lobes, and GRBs and their afterglows. 
Although turbulence in these environments may not necessarily be best described as being continuously driven by some external forcing (as opposed to being driven by an instability or freely decaying), and may not always achieve a stationary balance between turbulent heating and radiative cooling as envisioned in this paper, it is  nevertheless instructive to explore the implications of our model for these environments. 
In particular, here we shall consider the applications to the Crab Nebula and to blazar jets, while leaving the discussion of other types of astrophysical systems mentioned above to future studies.

\medskip
\noindent{\bf The Crab Nebula}\\
Based on broadband observations \cite[see, e.g.,][]{deJager_etal-1996, Aharonian_etal-2006, Hester-2008, Abdo_etal-2010}, the Crab Nebula is believed to be filled with mildly magnetized (with estimates for the magnetization $\sigma $ parameter ranging from $10^{-3}$ to~1) ultra-relativistic electron-positron pair plasma \citep{Rees_Gunn-1974, Kennel_Coroniti-1984a, Kennel_Coroniti-1984b}, see \cite{Arons-2012, Kargaltsev_etal-2015, Reynolds-2016} for recent reviews. It shines to us brightly across the entire observationally accessible electromagnetic spectrum, with most of the power attributed to synchrotron emission, spanning from radio to high-energy ($\sim 100$~MeV) gamma-rays, with a second, somewhat weaker (by about two orders of magnitude), distinct component at even higher (TeV) photon energies, attributed to inverse-Compton emission.  The presence of a very clear and prominent power law, extending over about 8 orders in photon energy, from optical (1~eV) to gamma-rays ($10^8$~eV), unambiguously indicates that the emitting plasma is non-thermal, characterized by a very broad power-law energy distribution of electrons and positrons. Detailed spectral modeling of the inner part of the Nebula yields characteristic magnetic field strengths of a few hundred micro-gauss and power-law particle spectra ranging from $10^5$ to~$10^9\, m_e c^2$. The roughly 1~eV peak of the synchrotron component then corresponds to particles with $\gamma\sim 10^6$ (and hence a characteristic energy $\gamma m_e c^2 \sim 0.5\, {\rm TeV}  \sim 1\, {\rm erg}$) dominating the emission.  The expected SSC photon energy, corresponding to the scattering of the synchrotron-peak 1~eV photons off of the same $\gamma \sim 10^6$ electrons, falls in the TeV band, consistent with the observed peak of the IC component. 

Although the plasma distribution in the Nebula is decidedly nonthermal, it is still interesting to examine the implications of the inferred $\gamma \sim 10^6$ dominant particle energy scale in light of our model. 
Substituting this value into equation~(\ref{eq-theta-tau_SSC}) leads to a Thomson optical depth of $\tau_T \sim 10^{-13}$, where we have conservatively taken $\kappa q_0 \beta_{\rm turb} \sim 0.1$. 
The latter estimate is motivated by the observations of plasma motions within the Nebula exhibiting speeds of a few tens of percent of light speed \cite[e.g.,][]{Hester-2008}.
Taking the driving-scale size in the inner part of the Nebula to be $L \sim 0.1 \, {\rm pc} = 3 \times 10^{17} \, {\rm cm}$, comparable to the size of the pulsar wind termination shock \citep{Rees_Gunn-1974}, leads then to an estimate for the density of energetically dominant ultra-relativistic particles of $n = \tau_T/\sigma_T L \sim 5\times 10^{-7} \, {\rm cm^{-3}}$, and hence a plasma internal energy density of 
$U_{\rm pl} = n \bar{\gamma} m_e c^2 \simeq n \cdot (0.8\, {\rm erg}) \sim 4\times 10^{-7} \, {\rm erg/cm^3}$.
This estimate, taken together with the observationally inferred typical magnetic field strength of $B \sim 300\, \mu {\rm G} = 3\times 10^{-4}\, {\rm G}$, and hence a magnetic energy density of $U_{\rm mag} = B^2/8\pi \sim 3.6 \times 10^{-9}\, {\rm erg/cm^3}$, implies a magnetization parameter of
\beq
\sigma = {B^2\over{4\pi \bar{\gamma} n m_e c^2}} = 2 U_{\rm mag} / U_{\rm pl}  \simeq 1.4 \times 10^{-2}\, ,
\eeq
which is broadly consistent with the expectations based on spectral modeling \cite[e.g.,][]{Kennel_Coroniti-1984a}.

Next, we can use equation~(\ref{eq-N}) to estimate the expected number and typical frequencies of the spectral  components.
Taking for convenience $B \simeq 4 \times 10^{-4}\, {\rm G}$, so that $b = B/B_Q \simeq 10^{-17}$ and $\tau_T \sim 10^{-13}$, we find $\hbar \Omega_c  = b\,m_e c^2 \simeq 5\times 10^{-12}\, {\rm eV}$, 
$\epsilon_{\rm synch}  \simeq \overline{\gamma^2} \, \hbar \Omega_c \sim 1\, {\rm eV}$,
$\epsilon_{\rm SSC}  \simeq \overline{\gamma^2} \epsilon_{\rm synch} \sim 1\, {\rm TeV}$,
$N \simeq 17/13 - 1/2 \simeq 0.8 \simeq 1$, so it is not surprising that there is only one observed SSC component. 
The SSC photons already have energies (1~TeV) comparable to that of the dominant electrons and positrons. 
The expected ratio between the SSC and synchrotron luminosities, $\mathcal{E}_{\rm SSC}/\mathcal{E}_{\rm synch} = r = \kappa q_0 \beta_{\rm turb}\sim 0.1$ is roughly consistent with, although by a factor of a few greater than, the observed ratio (of about~$10^{-2}$, see, e.g., \cite{Aharonian_etal-2004}).

%-----------------------------------------------------------------------------------------------------

\medskip 
\noindent{\bf Blazar jets}\\
Another natural application of the physical picture advanced in this paper is to blazar jets, namely to high-synchrotron-peaked BL Lacs objects (HBLs) populating the low-luminosity part of the blazar sequence \citep{Fossati_etal-1998}.
The high-energy IC emission component observed in the broadband spectra of these systems is comparable in power to the lower-energy synchrotron component and is commonly attributed to SSC emission, rather than to external IC as in higher-luminosity, low-synchrotron-peaked (LBL) class of blazars called Flat-Spectrum Radio Quasars (FSRQs) (see, e.g., \cite{Madejski_Sikora-2016} for a recent review). 

Theoretical models of flaring blazar jets, e.g., those developed to explain the ultra-rapid flares observed in HBLs at TeV energies by HESS, MAGIC, and VERITAS \cite[e.g.,][]{Aharonian_etal-2007, Albert_etal-2007,Acciari_etal-2011} imply typical comoving magnetic fields of order 10\,G 
and jet plasma (cold) ion magnetization parameters of order $\sigma = B^2 / 4 \pi n m_p c^2  \sim 100$ \citep{Giannios_etal-2009}.  
These values correspond to a comoving plasma density of 
$n = B^2/ 4 \pi \sigma m_p c^2 \sim 100\, {\rm cm^{-3}} \, B_1^2 \sigma_2^{-1}$, 
where $B_1\equiv B/10\,{\rm G}$ and $\sigma_2 \equiv \sigma/100$. 
This, in turn, indicates a Thomson optical depth of $ \tau_T = n \sigma_T L \sim 10^{-8} \, B_1^2 \sigma_2^{-1}\, L_{14}^{-1}$, where $L_{14} \equiv L/(10^{14}\, {\rm cm})$ is the emitting region's size normalized to a fiducial value of $10^{14}\, {\rm cm}$ inferred from blazar flare variability timescale \citep[e.g.,][]{Begelman_etal-2008, Ghisellini_Tavecchio-2008, Giannios_etal-2009} and comparable to or somewhat smaller than the gravitational radius of the central supermassive black hole powering the blazar (for $M_{\rm BH} \sim 10^9\, M_\odot$). 
Then,  equation~(\ref{eq-theta-tau_SSC}) readily yields an estimate for the typical Lorentz factor of the emitting relativistic electrons of $\gamma_e \sim \tau_T^{-1/2} \sim 10^4 \, B_1^{-1} \sigma_2^{1/2}\, L_{14}^{1/2}$. This is indeed consistent with the characteristic values inferred observationally from the ratio of the frequencies of the two emission peaks, typically observed to be in X-rays (keV) for synchrotron and very-high-energy (VHE) gamma-rays (100~GeV) for~IC \cite[e.g.,][]{Madejski_Sikora-2016}.
In addition, from the inferred typical values of $b\sim 10^{-13}$ and $\tau_T \sim 10^{-8}$, we expect, using equation~(\ref{eq-N}), that there should be only one SSC component, as is indeed observed. 
We thus conclude that the overall physical picture and the simple estimates presented in this paper describe the conditions in flaring HBL blazar jets reasonably well.

%-----------------------------------------------------------------------------------------------------

%***********************************************************************************

\section{Discussion and Conclusions}
\label{sec-conclusions}

In this study we have investigated the general thermodynamic and radiative properties of an optically thin MHD-turbulent relativistically hot plasma subject to strong radiative cooling in the steady-state regime where radiative cooling balances turbulent heating. In particular, we have explored the effects of synchrotron, synchrotron-self-Compton (SSC), and higher-order Compton cooling on setting the effective temperature (or the typical particle kinetic energy) of such a plasma. We have shown that, when turbulence is relativistic, i.e., when the rms fluid velocity at the large (driving) scale is comparable to the speed of light, the system tends to reach a statistical steady state characterized by an rms particle energy~$\gamma_{\rm rms} m_e c^2$ that is inversely proportional to the square root of the system's Thomson optical depth~$\tau_T$. 
This equilibrium state is a thermally stable attractor.
The  $\gamma_{\rm rms} \sim \tau_T^{-1/2}$ scaling is universal, essentially independent of the turbulent driving strength and hence of the corresponding magnetic field strength. The reason for this fundamental relationship between the temperature and the optical depth is that the MHD turbulent energy dissipation rate and the synchrotron cooling rate are both proportional to the magnetic energy density, which thus cancels out from the heating/cooling balance equation. We have then found that the plasma parameters inferred observationally in the Crab PWN and in HBL blazar jets are broadly consistent with this relationship. Furthermore, we have shown that, under a broad range of conditions, the SSC radiation power automatically becomes comparable (within a factor of order unity) to the synchrotron power. Moreover, when the normalized turbulent magnetic field, $b\equiv B/B_Q$, where $B_Q \simeq 4 \times 10^{13}\,{\rm G}$ is the critical quantum field, is much smaller than~$\tau_T$, one may expect several higher-order Compton radiation components, with roughly comparable power.  
The number of distinct IC components present in the system's spectrum is then controlled by Klein-Nishina effects and is approximately given by $N \simeq (\ln b /\ln \tau_T) - 1/2$. 
Finally, we have also derived the validity condition for synchrotron self-absorption to be negligible. 

%-----------------------------------------------------------------------------------------

We now would like to make a couple remarks about the applicability and possible extensions of our model. 
First, most astrophysical systems to which this analysis may in principle be relevant, including both of those discussed in the previous section, are not necessarily subject to continuous driving by some external force. Instead, it is more natural to think of them as expanding flows with turbulence either decaying or driven by some MHD-level, large-scale instabilities such as kink or Kelvin-Helmholtz. But even in these cases, the characteristic rate of turbulent energy dissipation can roughly be estimated by Eq.~\ref{eq-turb-heat}, with all the geometric and physical uncertainties absorbed into the dimensionless parameter~$\kappa$, which can still be expected to be of order~1. It is thus not surprising that the key qualitative predictions of the proposed theory should hold reasonably well in a broad class of systems.

Second, for our analysis to apply, the plasma does not need to be a pure pair plasma. One can also apply this model to the electron component of an electron-ion plasma in the semi-relativistic (ultra-relativistic electrons but sub-relativistic ions) or the fully relativistic regimes, such as may be expected in hot coronae and jets of accreting black holes. 
In this case the analysis is modified slightly because only a fraction~$f_e$, currently not very well constrained, of the total turbulent energy (which at the large, energy-containing scales is contained mostly in the bulk kinetic energy of the ions and the magnetic energy) gets transferred to the electrons and is ultimately radiated away. 
Most of the dissipated energy may thus go to ions and may not participate in the electron heating/cooling balance. 
It is then possible that one ends up with a two-temperature plasma, with $T_i \gg T_e$, expected, for example, in certain types of accretion flows onto black holes  \cite[e.g, ion tori, see][]{Rees_etal-1982, Narayan_Yi-1995, Quataert_Gruzinov-1999}. 
This consideration is especially important in situations where the plasma is collisionless, so that the electron-ion energy exchange due to Coulomb collisions is inefficient and hence $f_e$ is governed, instead, by complicated and still poorly understood collisionless plasma processes. While the electron heating fraction $f_e$ in relativistic plasma turbulence is currently not very well known \cite[see, however,][for nonrelativistic treatment]{Howes-2010}, it has recently been mapped out in PIC simulations of non-radiative relativistic electron-ion-plasma magnetic reconnection~\citep{Werner_etal-2018}.  
This and other recent theoretical and computational advances hold strong promise that we will have a full quantitative knowledge of $f_e$ as a function of various turbulence parameters within the next several years.

%{\bf Connection to Reconnection}
Second, we note that general arguments similar to those presented in this paper can also be explored, with some modifications, for relativistic magnetic reconnection with strong synchrotron cooling \cite[investigated, e.g., by][]{Lyubarsky-1996, Jaroschek_Hoshino-2009, Uzdensky_McKinney-2011, Uzdensky_Spitkovsky-2014, Uzdensky-2016, Beloborodov-2017}. 
In this case, the rate of energy dissipation per unit surface area of the reconnection layer is basically given by the Poynting flux into the layer and is therefore also proportional to~$B^2/8\pi$, with the coefficient of proportionality being the reconnection inflow velocity~$v_{\rm rec} = \beta_{\rm rec} c \sim 0.1 c$.  
The rate of synchrotron radiation losses per unit area is roughly $n\delta c \sigma_T\, (B^2/8\pi) \overline{\gamma^2}$, where $\delta$ is the effective thickness of the reconnection layer; it is thus also proportional to~$B^2/8\pi$. 
Then, upon balancing these two rates against each other and cancelling out~$B^2/8\pi$, we find 
$\overline{\gamma^2} \sim \beta_{\rm rec} \tau_{T,\delta}^{-1}$, where $\tau_{T,\delta} \equiv n \sigma_T \delta$ 
is the Thomson optical depth associated with the effective thickness $\delta$ of the layer \citep{Uzdensky_Spitkovsky-2014}.  

If we now make the usual assumptions that the plasma flow through the reconnection region is statistically stationary and approximately incompressible [which, however, may not be applicable in the case of reconnection with strong radiative cooling, see \cite{Uzdensky_McKinney-2011}], and also ignore pair production, then particle-number conservation relates the effective dimensionless reconnection rate~$\beta_{\rm rec}$ to the inverse aspect ratio $\delta/L$ of the layer (where $L$ is the layer's length): 
$\delta/L \sim v_{\rm rec}/v_{\rm out} \sim \beta_{\rm rec}$, 
where we assumed that the plasma outflow from the reconnection region is relativistic, $v_{\rm out} \sim c$. 
Then, one immediately obtains a relationship between the effective plasma temperature and the Thomson optical depth {\it along} the layer, $\tau_{T,L} \equiv n \sigma_T L$,  
\beq
\overline{\gamma^2} \sim {\beta_{\rm rec}\over{n \sigma_T \delta}} \sim {1\over{n \sigma_T L}} = \tau_{T,L}^{-1}\, , 
\eeq
which is essentially the same as the relationship~(\ref{eq-theta-tau_T}) that we have obtained for turbulence in the present paper.% 
\footnote{Similar arguments were presented by \cite{Uzdensky_McKinney-2011} for the case of non-relativistic reconnection with IC and cyclotron cooling and, more recently, by \cite{Beloborodov-2017} for relativistic reconnection with synchrotron and IC cooling and with pair creation.} 

%----------------------------------------------------------------------------------------

We hope that the simple physical model considered in this paper will help advance the emerging field of radiative plasma astrophysics by establishing a useful baseline and setting the stage for future, more elaborate theoretical and computational studies of radiative relativistic plasma turbulence, including first-principles radiative particle-in-cell simulations, and its application to high-energy astrophysics. 
In particular, we envision the following directions for future theoretical analyses and radiative PIC simulations: 
(1) investigating radiative plasma turbulence with strong synchrotron self-absorption effects; 
(2) incorporation of external IC cooling; 
(3) exploring the effects of nonthermal particle distributions; 
(4) investigating the electron heating fraction $f_e$ in semi-relativistic electron-ion plasmas; 
(5) exploring the effects of pair production; 
(6) generalizing the present steady-state picture to time-dependent scenarios, e.g., to freely expanding outflows (with $L \propto t$, $n\propto t^{-3}$); 
(7) investigating the effects of intermittency of energy dissipation, particle acceleration, and radiation emission.

%-----------------------------------------------------------------------------------------

\section*{Acknowledgements}

I would like to thank M.\,Begelman, A.\,Beloborodov, D.\,Lai, M.\,Medvedev, G.\,Werner, and V.\,Zhdankin for useful comments and encouraging and stimulating discussions. 
This work was supported by DOE grants DE-SC0008409 and DE-SC0008655, 
NASA grants NNX12AP17G, NNX16AB28G, and NNX17AK57G, and NSF grant AST-1411879.
I gratefully acknowledge the hospitality of the Institute for Advanced Study and the support from the Ambrose Monell Foundation.

%***********************************************************************************
%%%%%%%%%%%%%%%%%%%% REFERENCES %%%%%%%%%%%%%%%%%%

% The best way to enter references is to use BibTeX:

\bibliographystyle{mnras}
\bibliography{rad_turb}

% Alternatively you could enter them by hand, like this:
% This method is tedious and prone to error if you have lots of references
%\begin{thebibliography}{99}
%\bibitem[\protect\citeauthoryear{Author}{2012}]{Author2012}
%Author A.~N., 2013, Journal of Improbable Astronomy, 1, 1
%\bibitem[\protect\citeauthoryear{Others}{2013}]{Others2013}
%Others S., 2012, Journal of Interesting Stuff, 17, 198
%\end{thebibliography}

%***********************************************************************************

%%%%%%%%%%%%%%%%%%%%%%%%%%%%%%%%%%%%%%%%%%%%%%%%%%

%\end{document}

% Don't change these lines
\bsp	% typesetting comment
\label{lastpage}
\end{document}